\documentclass[twocolumn,showpacs,preprintnumbers,amsmath,amssymb]{revtex4}
\usepackage{epsfig}

\newcommand {\be}{\begin{equation}}
\newcommand {\ee}{\end{equation}}
\newcommand {\bea}{\begin{eqnarray}}
\newcommand {\ea}{\end{eqnarray*}}
\newcommand {\ba}{\begin{eqnarray*}}
\newcommand {\eea}{\end{eqnarray}}

\newcommand {\ham} {{\mathcal H}}

\newcommand {\ket}{\rangle}

\newcommand {\refeq}[1] {(\ref{#1})}
\newcommand{\bm}[1]{ \mbox{\boldmath $#1$}  }

\begin{document}
\title{Integral relations for three-body continuum states with the adiabatic expansion}

\author{P. Barletta,$^1$ C. Romero-Redondo,$^2$ A. Kievsky,$^3$ M. Viviani,$^3$ 
and E. Garrido$^2$}
\affiliation{$^1$Department of Physics and Astronomy, University College London,
 London WC1E 6BT, United Kingdom\\
$^2$Instituto de Estructura de la Materia, CSIC, Serrano 123, E-28006 Madrid, Spain \\
$^3$Istituto Nazionale di Fisica Nucleare, Largo Pontecorvo 3, 56100 Pisa, Italy}


\begin{abstract}
Application of the Hyperspherical Adiabatic expansion to describe
three-body scattering states suffers the problem of a very slow
convergence. Contrary to what happens for bound states,  a huge number
of hyperradial equations has to be solved, and even if done, the
extraction of the  scattering amplitude is problematic.  In this paper
we show how to obtain accurate scattering phase shifts  using the
Hyperspherical Adiabatic expansion.  To this aim two integral relations,
derived from the Kohn Variational  Principle, are used. The convergence
of this  procedure is as fast as for bound states.
\end{abstract}


\maketitle

\paragraph*{Introduction.---}
Few-body collisions
involving either nuclei, atoms, or molecules are frequently investigated.
To this aim different methods are at present available depending on the
interaction under study.
In nuclear physics, collisions involving three or four nucleons
have been extensively studied within the Faddeev method
or the Hyperspherical Harmonic (HH) method~\cite{faddeev,lazauskas,kievsky08}. 
These two
methods show sufficient flexibility to treat the complexities of the
nucleon-nucleon interaction. A different problem arises when the
interaction presents a hard core, as in the case of the atom-atom
interaction, or in systems with $A>4$. In the first case,
the Faddeev equations have been extended to deal with a hard core
repulsion~\cite{motovilov} whereas the Hyperspherical Adiabatic (HA)
expansion method proved to be a very efficient tool~\cite{nielsen01}.
In nuclear systems with $A>4$ tentatives to describe scattering states 
have recently appeared~\cite{nollett,sofia}.

Here we are interested in describing a 1+2 collision using
the HA expansion method. 
For bound states the convergence of the HA expansion
has proved to be very fast.
However the convergence of the expansion slows down significantly
in the case of low energy scattering states~\cite{barletta09}.
On the other hand
this method is extensively used to describe few-atom systems
in the ultracold regime (see Refs.~\cite{esry1,esry2} and references
therein) and, in particular,
atom-dimer collisions. Therefore a detailed study of its
convergence properties is timely.

In this letter it is shown for the first time how the HA expansion method
can be used to describe elastic scattering with a pattern of
convergence similar to a bound state calculation. This is achieved in
a simple but very general way in which a second order estimate of the
phase shift is extracted
from the wave function using two integral relations derived from the
Kohn Variational Principle (KVP)~\cite{kohn}. The number of HA
terms needed to obtain completely stable results
depends very little on the structure of the potential,
exactly as for bound state calculations.
The integral relations are governed by the wave function in the interaction
region. Therefore the stability of the results with a low number of HA basis
elements is a clear indication that inclusion of more terms in the expansion
only modifies the wave function outside the interaction range. 

As derived from the KVP, the integral relations are general and their application
is not limited to three particles. They can be applied to an $A$-body system
in which the scattering wave function is known in the interaction region.
Example of applications are given below.

\paragraph*{Continuum states in the Hyperspherical Adiabatic Expansion method.---}
The details of the HA method can be found in \cite{barletta09,nielsen01}.
For simplicity, here we restrict ourselves to three equal mass particles
with total angular momentum $L=0$ and with only $s$-waves involved.

   From the Jacobi coordinates
${\bm X}_i =  ({\bm r}_j - {\bm r}_k)/\sqrt{2}$ and
${\bm Y}_i =  ({\bm r}_j + {\bm r}_k - 2 {\bm r}_i)/\sqrt{6}$,
one defines the hyperspherical variables, 
$[\rho,\Omega_i]\equiv[\rho,\phi_i,\mu_i]$, with
$\mu_i={\hat {\bm X}}_i\cdot {\hat {\bm Y}}_i$, $X_i=\rho\cos\phi_i$,
and $Y_i=\rho\sin\phi_i$, where $\{i,j,k\}$ is a cyclic permutation
of $\{1,2,3\}$, and $\{ {\bm r}_i \}$ are the coordinates of the three
particles. In hyperspherical coordinates the Hamiltonian operator $\ham$
takes the form:
\be
\ham =  -\frac{\hbar^2}{2 m} T_\rho + \frac{\hbar^2}{2 m \rho^2}G^2
+ V(\rho,\Omega)
 =  -\frac{\hbar^2}{2 m} T_\rho + {\cal H}_\Omega    ,
\label{hhc}
\ee
where $T_\rho$ is the hyperradial operator, $G^2$ is the grand-angular
operator, $V(\rho,\Omega)=\sum_i V(X_i)$ is the potential energy,
and $m$ is set equal to the mass of the particles.
The wave function $\Psi$ for a specific bound or continuum state
is expanded as:
\be
\Psi^{ST\Pi} = \sum_{\nu=1}^{\infty} u_\nu(\rho) \Phi^{ST\Pi}_\nu(\rho,\Omega),
\label{adbasis}
\ee
where $S$, $T$, and $\Pi$ are the total spin, total isospin and parity.
For simplicity
we shall suppress from now on the corresponding labels in $\Psi$ and $\{\Phi_\nu\}$.
The HA basis elements $\{\Phi_\nu\}$ are the eigenfunctions of $\ham_\Omega$ at fixed
values of $\rho$. Their corresponding eigenvalues, $U_\nu(\rho)$, are the adiabatic
potentials, which enter in the coupled set of differential equations
(see Refs.~\cite{nielsen01,esry2})
\begin{eqnarray}
& \left[ -\frac{\hbar^2}{2 m}T_\rho +U_\nu(\rho)-\frac{\hbar^2}{2 m}Q_{\nu\nu}(\rho)-E
 \right] u_\nu(\rho)-  \nonumber \\
&\frac{\hbar^2}{2 m}{\displaystyle \sum_{\nu'\ne\nu}^{N_A}} 
 \left[ Q_{\nu\nu'}(\rho) +P_{\nu\nu'}(\rho)(\frac{5}{\rho}+2 \frac{d}{d\rho})
\right]u_{\nu'}(\rho)=0
\label{usys}
\end{eqnarray}
with $N_A$ the number of adiabatic channels included in the calculation,
$E$ the three-body energy, and from which the hyperradial
functions $u_\nu(\rho)$ are obtained.
At energies below the two-body breakup $E^{2B}$ and
$\rho\rightarrow\infty$, the total scattering wave function behaves
asymptotically as~\cite{barletta09}
\be
 \Psi \rightarrow 
 \phi_d(r)\left[ \frac{\sin{(k_\rho \rho)}}{\sqrt{k_\rho}\rho}
    + \tan\delta_\rho\frac{\cos{(k_\rho \rho)}}{\sqrt{k_\rho}\rho}\right]|ST\ket .
\label{eqas}
\ee
However, as we will show below, even increasing $N_A$ as much as possible,
the computed value of $\delta_\rho$ does not converge to the expected one.
This can be understood from the fact that the
asymptotic structure of the system can be constructed
in terms of the functions:
\be
\begin{aligned}
& F_{ST} = \sum_{i=1}^3 F_{ST}(i)=\sum_{i=1}^3 \phi_d(X_i)
\frac{\sin{[ k_y y_i]}}{\sqrt{k_y} \; y_i} |ST \ket  \cr
& G_{ST}= \sum_{i=1}^3 G_{ST}(i)=\sum_{i=1}^3 \phi_d(X_i)
\frac{\cos{[ k_y y_i]}}{\sqrt{k_y} \; y_i} |ST\ket  \;\; .
\label{ha1i}
\end{aligned}
\ee
where particle $i$ is assumed to hit the bound state made by $j$ and
$k$, and where $y_i=\frac{\sqrt{6}}{2}Y_i$ is the distance between
$i$ and the $j$-$k$ center of mass, and $k^2_y=\frac{2}{3}k_\rho^2$.
The asymptotic configuration in the limit $y_i\rightarrow\infty$ is then:
\begin{equation}
\Psi\longrightarrow F_{ST}+\tan\delta \; G_{ST}.
\label{eq8}
\end{equation}

When $\rho\rightarrow\infty$, the distance $X_i$ is limited by
$\phi_d$ and the approximate relation
$k_y y_i \approx k_\rho \rho$ holds. However, the exact equivalence
between $k_y y_i$ and $k_\rho \rho$ is not matched for
any finite value of $\rho$ and, accordingly, the boundary condition of
Eq.\refeq{eqas} is equivalent to the one in (\ref{eq8}) only at
$\rho\approx\infty$ and $N_A\rightarrow \infty$.
As a consequence $\delta_\rho$ converges extremely slowly to
$\delta$ by increasing the number of adiabatic states. This situation
has reduced the applicability of the method.

\paragraph*{Second order integral relations.---}

 From the above discussion 
and observing that in the expansion of the functions $F_{ST}$ and
$G_{ST}$ in terms of HA basis elements~\cite{barletta09}, the two
terms of Eq.~\refeq{eqas} represent the first term of that 
expansion, respectively,
the wave function $\Psi$ can be expressed asymptotically as:
\be
 \Psi= \sum^{N_A}_\nu u_\nu(\rho)\Phi_\nu(\rho,\Omega) \rightarrow
 AF_{ST}+ BG_{ST}  \;\;\; .
\label{eqas1}
\ee
In order to extract the coefficients $A$ and $B$ ($\tan\delta=B/A$), we
derive from the KVP two integral relations accurate up to second order.
The KVP states
that the following functional is stationary:
\begin{equation}
[\tan\delta]^{2^{nd}}=\tan\delta-<(1/A)\Psi|{\cal L}|(1/A)\Psi>
\label{kohn}
\end{equation}
 with respect to variations of the wave function, where
 ${\cal L}=\frac{2}{\sqrt{3}}\frac{m}{\hbar^2}({\cal H}-E)$. 
The scattering wave function can be schematically written as 
$(1/A)\Psi=\Psi_c+F_{ST}+\tan\delta \; {\widetilde G}_{ST}$.
The function ${\widetilde G}_{ST}$, representing a regularization of the 
function $G_{ST}$, introduces
 a nonlinear parameter
 $\gamma$  to eliminate a 
term proportional to $\delta({\bm y}_i)$ originated by $({\cal H}-E)G_{ST}$, 
and $\Psi_c$ is the part of the wave function 
inside the interaction region constructed in terms of 
some parameters (e.g. a linear combination of basis elements).
It verifies $\Psi_c\rightarrow 0$ asymptotically.
 The other
parameter in $\Psi$ is $\tan\delta$ . 
In the present work we have used
\begin{equation}
{\widetilde G}_{ST}= \sum_i \phi_d(X_i)
\frac{\cos{[ k_y y_i]}}{\sqrt{k_y} \; y_i} (1-{\rm e}^{-\gamma y_i})|ST\ket .
\end{equation}
The variation of the functional \refeq{kohn} with respect to the parameters in 
$\Psi_c$ and with respect to $\tan\delta$ leads to: 
\begin{equation}
<\Psi_c|{\cal L}|\Psi>=0 \hspace{0.35cm};
<{\widetilde G}_{ST}|{\cal L}|\Psi>=0   .
\label{psic}
\end{equation}
These two equations can be interpreted in two different ways.
In the case in which $\Psi$ is explicitly separated in the three 
terms $\Psi_c,
F_{ST},{\widetilde G}_{ST}$, the above equations are used to
determine $\Psi_c$ and the first order estimate of $\tan\delta$
($\tan\delta^{1^{st}}$). Accordingly, $\Psi$ is constructed after
solving these equations. A different case arises when $\Psi$ is
known (for example using the HA expansion) but the
separation in the three terms is not explicitly known. For this case
the two equations can be used to define of $\Psi_c$
and $\tan\delta^{1^{st}}$.

Introducing Eq.~\refeq{psic} into the functional,
the second order estimate of $\tan\delta$ is obtained:
\begin{equation}
[\tan\delta]^{2^{nd}}=(\tan\delta)^{1^{st}}-<F_{ST}|{\cal L}|(1/A)\Psi> 
\label{kohn1}
\end{equation}
with $A=<\Psi|{\cal L}|{\widetilde G}_{ST}>$. This is a
consequence of the general relation
$A=<\Psi|{\cal L}|{\widetilde G}_{ST}>-<{\widetilde G}_{ST}|{\cal L}|\Psi>$
(obtained by transforming the Laplacian term in a surface integral),
the normalization relation 
$<F_{ST}|{\cal L}|{\widetilde G}_{ST}>-<{\widetilde G}_{ST}|{\cal L}|F_{ST}> =1$
and the last equation derived from the KVP in Eq.~\refeq{psic}.
The same relation can be used to obtain
a first order estimate for the coefficient $B$ as
\begin{equation}
 B^{1^{st}}= <F_{ST}|{\cal L}|\Psi>-<\Psi|{\cal L}|F_{ST}>  .
\label{first}
\end{equation}
After multiplying Eq.~\refeq{kohn1} by $A$ one gets
that $B^{2^{nd}}=B^{1^{st}}-<F_{ST}|{\cal L}|\Psi>$, which
by use of Eq.~\refeq{first} leads to 
a second order integral relation for $B$
and, accordingly, a second order estimate for $\tan\delta$.
These results are the main conclusions of this paper and 
can be summarized as
\begin{equation}
\left.
\begin{array}{ccc}
B^{2^{nd}}&=&-<\Psi|{\cal L}|F_{ST}> \\
A&=& <\Psi|{\cal L}|{\widetilde G}_{ST}>
\end{array} \right\}
\tan\delta^{2^{nd}}=B^{2^{nd}}/A.
\label{second}
\end{equation}
The relations of Eq.\refeq{second}
are equivalent to the KVP and are useful in the cases in which
$\Psi$ is known but its explicit asymptotic form in terms of the functions
$F_{ST}$ and $G_{ST}$ is not. This is the case, for example,
when $\Psi$ is obtained from the solution of the HA equations. 
The integrands in the integral relations of Eq.~\refeq{second} go rapidly
to zero as $\;\rho\rightarrow\infty$  since
$F_{ST}$ and ${\widetilde G}_{ST}$ are solutions of 
${\cal L}$ in that limit. Therefore
an accurate knowledge of $\Psi$ outside the range of interaction is not needed.
In the present case, the explicit form of the integrals in Eq.(\ref{second}) are:
\begin{equation}
\begin{aligned}
&B^{2^{nd}}=-C
\int d\rho \rho^5 d\Omega \Psi(\rho,\Omega)V(X_i)[F_{ST}(j)+F_{ST}(k)] \cr
&A=C
\int d\rho \rho^5 d\Omega \Psi(\rho,\Omega)V(X_i)
[ {G}_{ST}(j)+{G}_{ST}(k)] +I_\gamma
\end{aligned}
\nonumber
\label{explicit}
\end{equation}
where $C=2\sqrt{3}m/\hbar^2$ and
$I_\gamma$ is a (short-range) integral including all terms depending on $\gamma$. 
Let us note that the
last integral is largely independent of $\gamma$ 
provided that the regularization is performed inside the interaction region
and $\Psi$ tends to the exact wave function.
The dependence of $\tan\delta$ on $\gamma$ is studied below.
We have found that values of 
$\gamma\approx\sqrt{m|E^{2B}|/\hbar^2}$ are a convenient choice.

The integral relations, as given above, are a generalization of the relation
used in the two-body case \cite{holt72}. 
However, an attempt to identify
the single term $B^{2^{nd}}=\tan \delta_B$ as a corrected phase-shift fails 
as we show below. 
The validity of Eq.~\refeq{second} is not limited to the
three-body case and to wave functions obtained using the HA method.
It can be applied to any wave function $\Psi$ which verifies
$({\cal H}-E)\Psi=0$ in the interaction region without any explicit
indication of its asymptotic behaviour. An example is represented by
the solution of $({\cal H}-E)\Psi=0$ in a box in which $\Psi$ is
set to zero at some distance. Using Eq.~\refeq{second} a second order
estimate of a phase shift can be obtained studying its convergence in terms
of the dimension of the box.

\paragraph*{Results.---}

As a first application we consider a three-body system of identical spinless
bosons interacting through a central, $s$-wave, gaussian potential 
$V(r)=V_0 \exp{-(r/r_0)^2}$, with $V_0=-51.5$ MeV and $r_0=1.6$ fm. Though this
potential is unrealistic, it will serve to the purpose of testing the method
due to the very extended dimer wave function 
($E^{2B}=0.397743$ MeV and $\hbar^2/m=41.4696$ ${\rm MeV}\;{\rm fm}^2$).
Such three-body system has two $L$=0 bound states with separation energies
$E_{3B}^{(0)}=-9.7574$ MeV and $E_{3B}^{(1)}=-0.4816$ MeV, respectively. 
In Table \ref{tab1} we show
the convergence of these two states in terms of the HH and HA expansions. From 
the table we observe the fast convergence of the HA even
in the case of the shallow state $E_{3B}^{(1)}$. We can conclude that 10
HA basis states are sufficient to describe simultaneously both bound
states.

\begin{table}[hbt]
\begin{tabular}{ccccccc}
  & \multicolumn{2}{c}{$E_{3B}^{(0)}$} & \hspace{2cm}& 
  & \multicolumn{2}{c}{$E_{3B}^{(1)}$} 
\\ \cline{2-3} \cline{6-7}
$N$ & HH & HA  &\hspace{1cm} & $N$& HH & HA  \\
1  & -9.2062 & -9.7347 & &  1 &     -    &  -0.4781  \\
2  & -9.5810 & -9.7552 & &  4 &     -    &  -0.4815  \\
3  & -9.7247 & -9.7573 & & 10 &  -0.2323 &  -0.4816  \\
4  & -9.7424 & -9.7574 & & 30 &  -0.4635 &  -0.4816  \\
6  & -9.7558 & -9.7574 & & 50 &  -0.4790 &  -0.4816  \\
8  & -9.7571 & -9.7574 & & 70 &  -0.4811 &  -0.4816  \\
10 & -9.7574 & -9.7574 & &100 &  -0.4815 &  -0.4816  \\
\end{tabular}
\caption{
Convergence of the bound state energies (in MeV) 
as a function of the number $N$ of HH and HA basis functions.}
\label{tab1}
\end{table}

We now show results for the $L$=0 phase shift at $E=-0.1$ MeV. 
Eq.\refeq{usys} has been solved up to $\rho$=500 fm with the boundary condition 
of Eq.\refeq{eqas} for increasing values of $N_A$. Due to the large 
extension of the $P_{\nu\nu'}$ and $Q_{\nu\nu'}$ coupling terms, the 
higher $N_A$ the larger the 
value of $\rho$ at which the asymptotic form of $u_1$ is verified.
For example, using only one HA basis state (only one hyperradial equation 
has to be solved), $u_1$ reaches its asymptotic form at $\rho\approx  100$ fm. When 
$40$ HA terms are used, this happens beyond 500 fm. The results
for $\delta_\rho$, $\delta_B$=$\arctan(B^{2^{nd}})$, $\delta^{2^{nd}}$=
$\arctan(B^{2^{nd}}/A)$ (for different values of $\gamma$), and $A$,
are given in Table~\ref{tab2} up to $40$ HA basis functions.

\begin{table}[hbt]
\begin{tabular}{c|c|cc|cccc}
$N_A$ & $\delta_\rho$ &$\delta_B$& $A$ & \multicolumn{4}{c}
{$\delta^{2^{nd}}=\arctan(B^{2^{nd}}/A)$}  \\
\hline
   &       & && $\gamma=0.1$ &$\gamma=0.25$&$\gamma=0.5$&$\gamma=1.0$ \\
\hline
1  & 65.23 & 67.85 &0.607& 75.895 & 76.212 & 75.479 & 74.727 \cr
4  & 71.65 & 71.89 &0.909& 73.385 & 73.446 & 73.450 & 73.429 \cr
8  & 72.32 & 72.42 &0.951& 73.222 & 73.237 & 73.245 & 73.259 \cr
12 & 72.56 & 72.62 &0.965& 73.190 & 73.194 & 73.196 & 73.203 \cr
16 & 72.67 & 72.71 &0.971& 73.182 & 73.183 & 73.185 & 73.189 \cr
20 & 72.72 & 72.76 &0.974& 73.180 & 73.180 & 73.182 & 73.186 \cr
24 & 72.75 & 72.79 &0.976& 73.179 & 73.179 & 73.180 & 73.184 \cr
28 & 72.77 & 72.81 &0.977& 73.179 & 73.179 & 73.180 & 73.184 \cr
32 & 72.78 & 72.82 &0.978& 73.179 & 73.179 & 73.180 & 73.184 \cr
36 & 72.79 & 72.82 &0.978& 73.179 & 73.179 & 73.180 & 73.183 \cr
40 & 72.79 & 72.83 &0.978& 73.179 & 73.179 & 73.180 & 73.183 \cr
\hline
\end{tabular}
\caption{Patterns of convergence for $\delta_\rho$, $\delta_B=\arctan{B}$, 
 $\delta^{2^{nd}}$ (in degrees) and $A$,
in terms of $N_A$ for E=$-0.1$ MeV.
The values of $A$ have been calculated using $\gamma=0.25$.}
\label{tab2}
\end{table}

\begin{figure}[hbt]
\epsfig{file=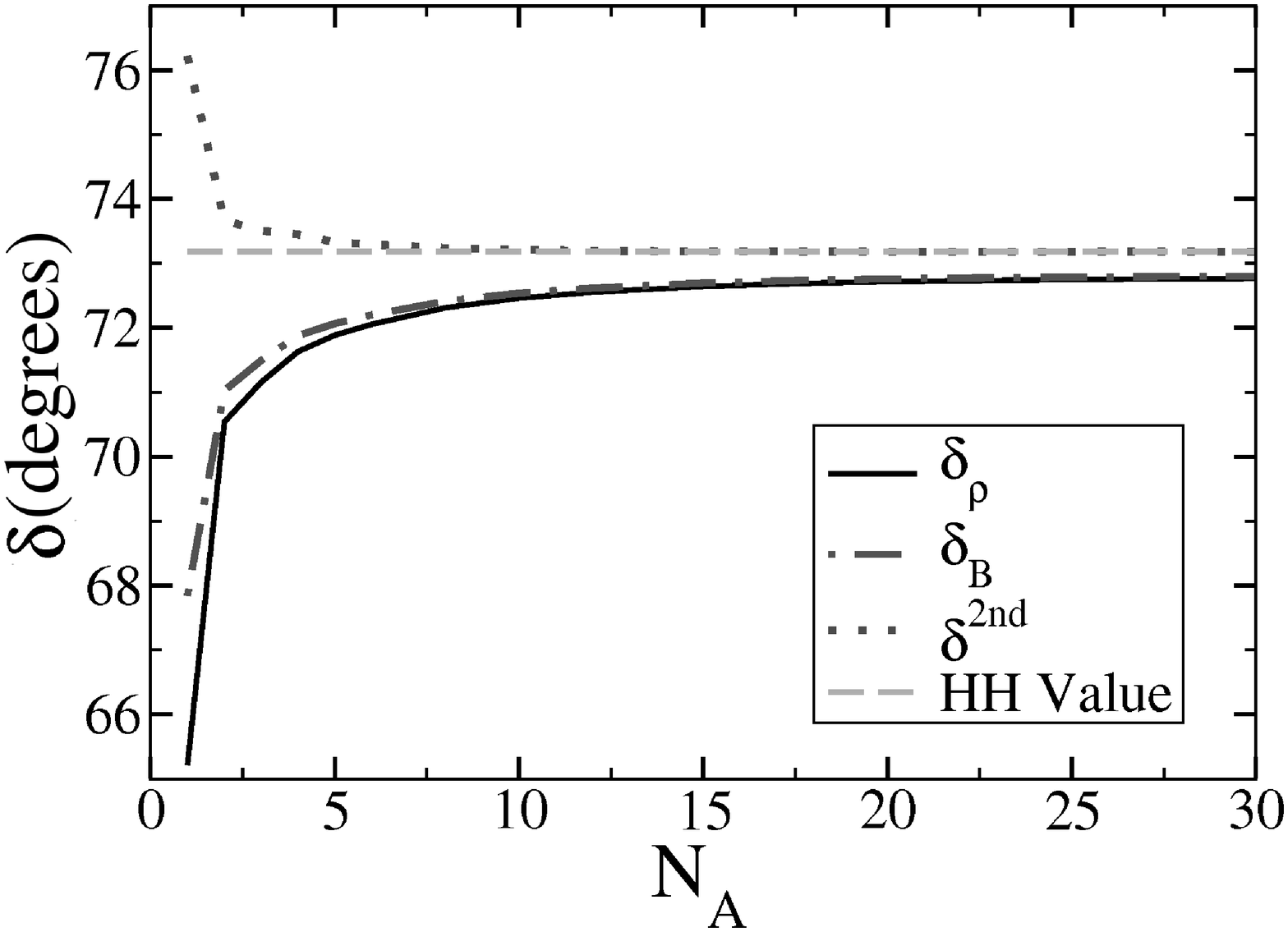, width=9cm, angle=0}
\vspace*{-8mm}
\caption{Convergence of $\delta_\rho$, $\delta_B$, and $\delta^{2^{nd}}$ 
as a function
of the number $N_A$ of HA basis functions. The converged value obtained from
the HH expansion is shown for comparison. 
}
\label{fig1}
\end{figure}

We observe that $\delta_\rho$ 
and $\delta_B$ converge very slowly to a value that, 
by extrapolation, can be estimated in the interval
$72.8^\circ - 72.9^\circ$ as $N_A\rightarrow\infty$.
This is at variance with the value 
$\delta^{2^{nd}}=73.18^\circ$, which shows a rate of convergence extremely
fast and a large stability with $\gamma$, as $N_A$ increases.
The calculation of $\delta^{2^{nd}}$ 
requires the knowledge of the radial functions up to values of 
$\rho$ not larger than $70-80$
fm, for which a relatively small number of HA terms is enough. Conversely,
$\delta_\rho$ and $\delta_B$ would converge to the correct phase shift only after 
imposing the 
boundary condition to the wave function at $\rho=\infty$, for which in principle 
infinitely many HA basis terms are needed.  
For comparison, a converged value of $73.180^\circ$ is 
obtained for the phase-shift using the HH expansion method with 120 basis elements. 
This result is in complete agreement with the one
obtained with Eq.(\ref{second}). 
The different patterns of convergence can be clearly seen on Fig.\ref{fig1}.

\begin{table}[hbt]
\begin{tabular}{cccc}
$N_A$ &  0.2 MeV    &  1.0 MeV  &   2.0 MeV  \cr
\hline
  4 &   -28.277   &  -55.875  &   -71.507  \cr
  8 &   -28.290   &  -55.865  &   -71.475  \cr
 12 &   -28.293   &  -55.864  &   -71.473  \cr
 16 &   -28.294   &  -55.863  &   -71.473  \cr
 20 &   -28.294   &  -55.863  &   -71.473  \cr
\hline
HH &-28.294 &-55.863 &-71.474  \cr
\hline
\end{tabular}
\caption{Convergence of $\delta^{2^{nd}}=\arctan(B^{2^{nd}}/A)$
(in degrees) at three incident energies with the MT-III potential.}
\label{tab3}
\end{table}

Ref.~\cite{barletta09} reports calculations at three different
energies using the MT-III potential~\cite{mtiii}, which has a yukawian repulsion 
at short distances, in the $S=3/2,T=0$ state. 
When the HH expansion is used, more than 120 basis states have 
to be included to reach convergence in the phase shifts. In Table~\ref{tab3}
we show the corresponding results when using Eq.(\ref{second}). 
In the last row the results from Ref.~\cite{barletta09} using the HH expansion
are given for comparison. 
The results obtained with the integral relations are
in complete agreement with those obtained using the HH expansion and show
a very fast pattern of convergence.

\paragraph*{Conclusions.---}
We have
derived two integral relations from the KVP which are accurate up to
second order. Their ratio, the phase shift, converges in terms of the HA
basis elements as fast as the binding energy in a bound state
calculation. The fast convergence has been shown for different
types of interactions. We would like to stress the general
validity of Eq.~\refeq{second}. Its application
will be very useful in the case in which
$\Psi$ is known in the interaction region
but the exact construction of its asymptotic form  is difficult.
The HA expansion has been applied in Refs.~\cite{esry2,incao} to
compute phase-shifts in a 1+2 and a 2+2 helium atom collisions. Accordingly.
Eq.~\refeq{second} can be used directly to obtain a second order estimate
of the phase-shifts. In Ref.~\cite{nollett}, $n-\alpha$ scattering has been
studied using Quantum Monte Carlo techniques. The wave function of the system
was obtained solving $({\cal H}-E)\Psi=0$ in a box. The knowledge of $\Psi$
in the interaction region allows for a direct application of Eq.~\refeq{second}
also in this case.
To be noticed that in the case in which more than one elastic channel is open,
the coefficients $A$ and $B$ of Eq.~\refeq{second}
correspond to matrices 
\begin{equation}
\left.
\begin{array}{ccc}
B^{2^{nd}}_{ij}&=&-<\Psi_i|{\cal L}|F_j> \\
A_{ij}&=& <\Psi_i|{\cal L}|{\widetilde G}_j>
\end{array} \right\}
R^{2^{nd}}=A^{-1}B^{2^{nd}}.
\label{second1}
\end{equation}
with $R^{2^{nd}}$ the second order estimate of the scattering matrix 
whose eigenvalues are the phase shifts and the indeces $(i,j)$ indicate
the different asymptotic configurations accessible at the specific energy
under consideration. Finally we would like to mention the possibility of
using Eq.(13) to describe a 1+2 elastic collision with charged particles
using a screened Coulomb potential and free asymptotic conditions.

\paragraph*{Acknowledgments---}This work was partly supported by DGI of MEC (Spain),
contract No.  FIS2008-01301. One of us (C.R.R.) thanks
support by a predoctoral I3P grant from CSIC and the European
Social Fund.

\end{document}